\documentclass[times,twocolumn,final]{elsarticle}
\usepackage{framed,multirow}
\usepackage{amsmath}
\usepackage{amssymb}
\usepackage{latexsym}
\usepackage{url}
\usepackage{xcolor}
\usepackage{hyperref}
\usepackage{booktabs}
\usepackage{color}
\usepackage{geometry}
\graphicspath{{./Figs/}}

\def\x{{\bf x}}

\def\0{{\bf 0}}
\def\1{{\bf 1}}

\def\etal{{\em et al.}}
\def\eg{{\em e.g.}}
\def\ie{{\em i.e.}}

\def\etal{{\em et al.\/}\,}

\definecolor{newcolor}{rgb}{.8,.349,.1}

\begin{document}
\begin{frontmatter} 
\title{Brain Anatomy Prior Modeling to Forecast Clinical Progression of Cognitive Impairment with Structural MRI\\
} 
\author{Lintao Zhang, Jinjian Wu, Lihong Wang, Li Wang, David C. Steffens, Shijun Qiu, Guy G. Potter\corref{cor1}}
\author{Mingxia Liu\corref{cor1}}

\cortext[cor1]{Corresponding authors:  G.~Potter (guy.potter@duke.edu) and M.~Liu (mxliu@med.unc.edu).}

\begin{abstract}
Brain structural MRI has been widely used to assess the future progression of  cognitive impairment (CI).  
Previous learning-based studies usually suffer from the issue of small-sized labeled training data, while there exist a huge amount of structural MRIs in large-scale public databases. 
Intuitively, brain anatomical structures derived from these public MRIs (even without task-specific label information) can be used to boost CI progression trajectory prediction. 
However, previous studies seldom take advantage of such brain anatomy prior. 
To this end, this paper proposes a brain anatomy prior modeling (BAPM) framework to forecast the clinical progression of cognitive impairment with small-sized target MRIs by exploring anatomical brain structures. 
Specifically, the BAPM consists of a \emph{pretext model} and a \emph{downstream model}, with a shared brain anatomy-guided encoder to model brain anatomy prior explicitly. 
Besides the encoder, the pretext model also contains two decoders for two auxiliary tasks (\ie, MRI reconstruction and brain tissue segmentation), while the downstream model relies on a predictor for classification.
The brain anatomy-guided encoder is pre-trained with the pretext model on 9,344 auxiliary MRIs without diagnostic labels for anatomy prior modeling. 
With this encoder frozen, the downstream model is then fine-tuned on limited target MRIs for prediction. 
We validate the BAPM on two CI-related studies with T1-weighted MRIs from 448 subjects.  
Experimental results suggest the effectiveness of BAPM in (1) four CI progression prediction tasks, (2) MR image reconstruction, and (3) brain tissue segmentation, compared with several state-of-the-art methods. 

\end{abstract}
\end{frontmatter}

\section{Introduction}
\label{S1}
{S}tructural brain anatomy information provided by magnetic resonance imaging (MRI) has been increasingly used to forecast clinical progression of cognitive impairment (CI) in various clinical and research fields~\cite{ashtari2022multi, el2021personalized, gonuguntla2022brain, guo2020novel, lombardi2020structural, yin2023anatomically}. 
There are many learning-based methods developed for MRI-based CI progression prediction, but typically rely on large amounts of labeled brain MRI scans for model training, especially for data-greedy deep learning approaches.  
Unfortunately, it is very challenging to acquire MRIs with diagnostic labels in clinical practice~\cite{nanni2020comparison}. 
On the other hand, there are a large number of brain MRIs in public datasets such as the Alzheimer's Disease Neuroimaging Initiative (ADNI)~\cite{jack2008alzheimer}. 
Even without task-specific label information, these public MRI data can intuitively provide rich brain anatomy prior, while such prior knowledge can be potentially employed to improve the performance of deep learning models for CI progression prediction.

Several previous studies have proposed different strategies to model brain anatomy priors to boost learning performance. 
For instance, Song~\etal~\cite{song2016anatomy} utilize tumor boundary contrast between fluid-attenuated inversion recovery (FLAIR) MRI and T2-weighted MRI for tumor segmentation, while Yamanakkanavar~\etal~\cite{yamanakkanavar2020mri} show that brain MRI segmentation priors can improve model's diagnosis performance. 
Recently, some researchers have proposed several deep learning models that rely on medical image reconstruction without requiring specific category labels, which can be pre-trained on unlabeled MRIs from existing large-scale datasets.
For instance, several studies~\cite{zhou2021models, zhou2021preservational} use CT image reconstruction as a pre-training task to train a deep model that can be transferred to classification and segmentation tasks. 
However, these studies generally require that source and target domains share identical label distributions and few studies utilize such brain anatomy priors to assess the clinical progression of CI with MRI. 
\begin{figure*}[!t]
\centering
\includegraphics[width=0.99\textwidth]{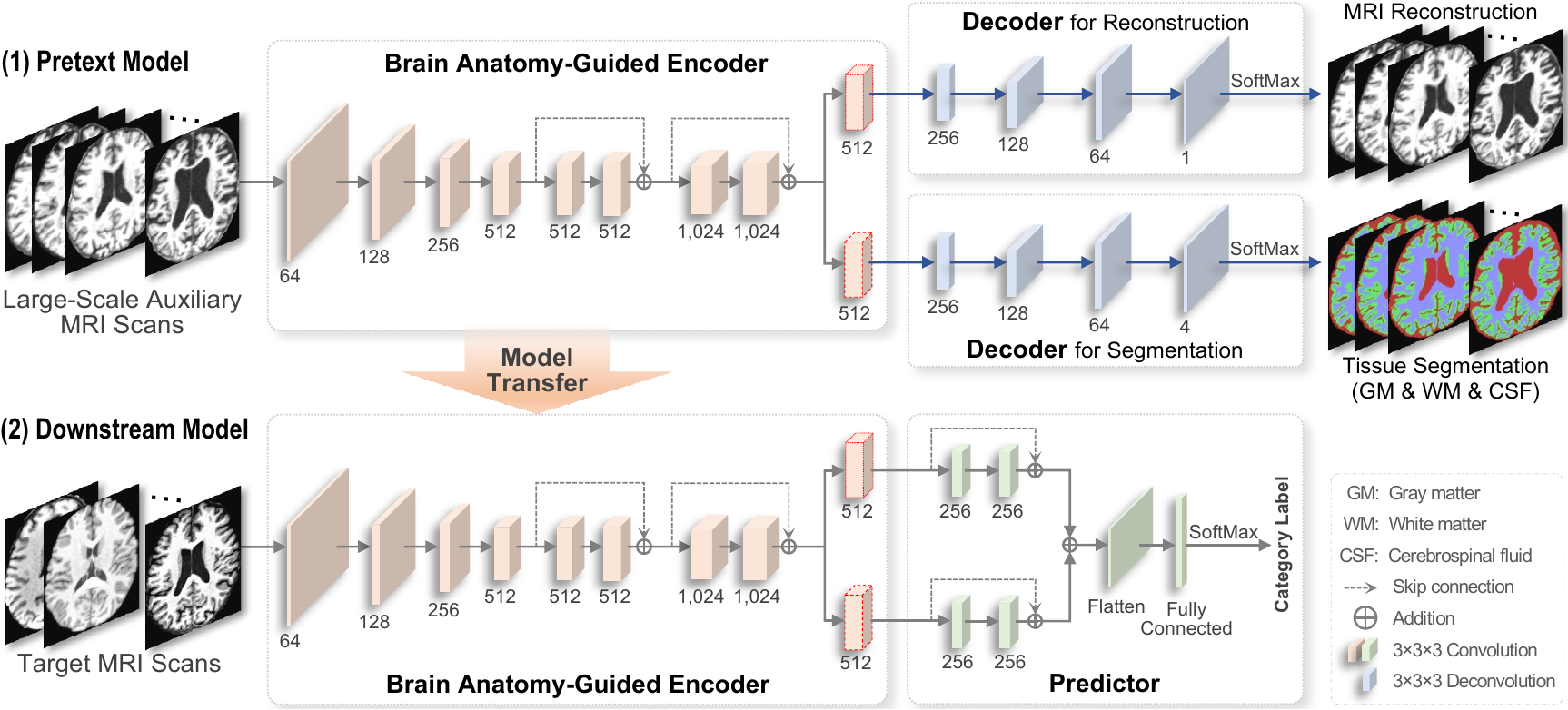}
\caption{Illustration of the proposed brain anatomy prior modeling (BAPM) framework for forecasting clinical progression of cognitive impairment.  
The BAPM consists of (1) a \emph{pretext model} and (2) a \emph{downstream model}, with a shared brain anatomy-guided encoder for brain anatomy prior modeling. 
The pretext model also contains two decoders for auxiliary tasks (\ie, MRI reconstruction and brain tissue segmentation), while the downstream model relies on a predictor for disease identification.
The brain anatomy-guided encoder is trained with the pretext model on 9,344 T1-weighted MRIs from ADNI~\cite{jack2008alzheimer} for brain anatomy prior learning. 
The downstream model is then fine-tuned on target MRIs with this learned encoder frozen.
}
\label{figFramework}
\end{figure*}
To address the issue, we propose a brain anatomy prior modeling (BAPM) framework to predict clinical progression of cognitive impairment with T1-weighted MRIs, incorporated with brain anatomy priors provided by both MRI reconstruction and brain tissue segmentation auxiliary tasks. 
As shown in Fig.~\ref{figFramework}, the BAPM consists of a \emph{pretext model} and a \emph{downstream model} that share a brain anatomy-guided encoder for brain anatomy prior modeling. 
Besides the encoder, the pretext model also contains two decoders for image reconstruction and tissue segmentation, respectively, providing guidance information for model training based on large-scale auxiliary source MRIs. 
Besides, the downstream model relies on a predictor for disease diagnosis and/or prognosis on target MRI data. 
In BAPM, we first train the pretext model on 9,344 MRI scans from ADNI without any category labels, and then transfer the encoder to the downstream model for fine-tuning and prediction on target MRIs.   
To the best of our knowledge, this is among the first attempts that utilize brain anatomy prior derived from large-scale public MRIs to assess the clinical progression of cognitive impairment. 
The source code has been released to the public via GitHub\footnote{https://github.com/goodaycoder/BAPM}.

The main contributions of this work are summarized below.
\begin{itemize}
    \item A brain anatomy prior modeling (BAPM) framework is developed to forecast clinical progression of cognitive impairment with T1-weighted MRIs, where both MRI reconstruction and brain tissue segmentation are used as auxiliary tasks to explicitly model brain anatomy priors. 
    \vspace{-2mm}
    \item
    For source MRIs used in the pretext model, the MRI reconstruction task does not require any label information, while the brain tissue segmentation task relies on ground-truth segmentation maps. 
    In contrast, category labels (\eg, CI) of target MRIs are required for downstream model fine-tuning. 
    That is, source and target MRIs used in our BAPM can have different label distributions, thus significantly improving its practical utility. 
    \vspace{-2mm}
    \item The pretext model of BAPM is trained on large-scale (\ie, $>9,000$) T1-weighted MRIs from the public ADNI dataset. Besides the encoder that is further transferred to the downstream prediction task, our trained decoders can be straightforwardly applied to other tasks of image reconstruction and brain tissue segmentation. 
    \vspace{-2mm}
    \item Extensive experiments have been performed on two CI-related studies in four classification/prediction tasks, with quantitative and qualitative results suggesting the effectiveness of BAPM incorporated by brain anatomy priors. 
\end{itemize}

The remainder of this paper is organized as follows. 
Section~\ref{S2} reviews the most relevant studies. 
Section~\ref{S3} introduces the materials and proposed framework. 
In Section~\ref{S4}, we introduce experimental setup, competing methods and experimental results. 
Section~\ref{S5} analyzes influences of several key components of BAPM and discusses the limitations of the current work as well as future research directions. 
Finally, this paper is concluded in Section~\ref{S6}.

\begin{table*}[!tbp]
\renewcommand\arraystretch{1}
\centering
\setlength{\belowdisplayskip}{0pt}
\setlength{\abovedisplayskip}{0pt}
\setlength{\abovecaptionskip}{0pt}
\setlength{\belowcaptionskip}{0pt}
\footnotesize
\caption{Diagnostic and demographic information and category labels of studied subjects in two cognitive impairment related studies. The values are denoted as ``mean$\pm$standard deviation''. F/M: Female/Male, MMSE: Mini-Mental State Examination. Demographic information of subjects from the late-life depression (LLD) study at baseline time was reported, while their diagnostic labels were determined based on 5-year follow-up diagnosis.}
\setlength{\tabcolsep}{1.2mm}{}{
\begin{tabular}{l|l|l |c c c c c}   
\toprule
{Study} & {Category} &{Description}  &{Gender (F/M)} & {Age} &{Education (Years)} & MMSE\\
\midrule 
\multirow{3}{*}{Late-Life Depression (LLD)} &CN &Cognitively normal & 59/30   &70.2$\pm$6.8 &15.6$\pm$1.9  &29.1$\pm$1.3\\ 

                    &CND &Cognitively normal with depression  & 120/59  &67.6$\pm$6.5 &15.5$\pm$2.3  &28.9$\pm$1.3 \\
                    &CID  &Cognitive impairment or dementias with depression & 22/19    &73.7$\pm$6.3 &15.1$\pm$2.6  &27.5$\pm$3.4 \\
\midrule 
\multirow{3}{*}{Diabetes Mellitus (DM)} &CN &Cognitively normal & 24/21   &47.8$\pm$8.5 &11.4$\pm$3.1  &29.0$\pm$1.0 \\
                    &DM &Cognitively normal with diabetes mellitus & 28/29    &45.5$\pm$8.2 &12.8$\pm$3.1 &28.7$\pm$1.3 \\
                    &MCI &Mild cognitive impairment with diabetes mellitus & 14/23    &51.8$\pm$9.3 &10.2$\pm$4.5  &26.9$\pm$2.8 \\
\bottomrule
\end{tabular} 
\label{Diagnosis_information}
}
\if false
\begin{tablenotes}
    \item[1] * category labels are determined based on a 5-year follow-up diagnosis.
\end{tablenotes}
\fi 

\end{table*}
\section{Related Work}
\label{S2}
\subsection{Cognitive Impairment Progression Analysis with MRI}
Many previous MRI-based studies have shown that cognitive impairment and relevant brain diseases are related to structural brain changes~\cite{feng2020mri, manschot2006brain, rosenberg2019magnetic, joseph2021structural, takamiya2021lower}.
For instance, Manschot~\etal~\cite{manschot2006brain} conclude that cognitive impairments of type 2 diabetes subjects are associated with brain subcortical ischemic changes and increased brain atrophy. 
In research related to CI assessment, traditional methods mainly use statistical analysis~\cite{willette2014prognostic} and machine learning~\cite{rallabandi2020automatic, lebedeva2017mri, yao2018ensemble} methods to explore potential relationships between brain anatomy features and brain diseases.
These methods usually rely on preprocessing tools (\eg, FSL~\cite{jenkinson2012fsl}, FreeSurfer~\cite{fischl2012FreeSurfer}, and SPM~\cite{ashburner2012spm}) to extract brain anatomical features from MRI scans for analysis.
For instance, Hedges~\etal~\cite{hedges2022reliability} extract MRI features (such as cortical/subcortical volumes, cortical surface areas, and cortical thickness) using FreeSurfer in a longitudinal study. 
In recent years, research using deep learning for automatic MRI feature extraction has become increasingly popular and achieved promising performance in MRI-based brain disorder analysis~\cite{noor2019detecting, frizzell2022artificial, basaia2019automated}.
Some studies explore the link between changes in brain structure and cognitive decline~\cite{yin2023anatomically} or use brain anatomy priors to help improve deep models' performance~\cite{song2016anatomy, yamanakkanavar2020mri}.
For example, Yin~\etal~\cite{yin2023anatomically} reveal that there exist aging-related neuroanatomy differences between normal controls and patients with Alzheimer's disease (AD) and mild cognitive impairment.  
Intuitively, modeling such anatomy knowledge in brain MRI can promote learning performance. 
In this work, we propose to \emph{explicitly model brain anatomy priors} from T1-weighted MRIs through two auxiliary tasks, including MRI reconstruction and brain tissue segmentation.

\subsection{Learning with Limited Neuroimaging Data}
Existing learning-based methods typically rely on large amounts of labeled brain MRI scans for model training, but it is very challenging to get diagnostic labels in clinical practice~\cite{nanni2020comparison}. 
Many strategies have been developed to handle small-sample-size issues in neuroimage analysis by leveraging large-scale auxiliary source data, such as domain adaptation and transfer learning techniques. 
As a popular solution, domain adaptation aims to reduce distribution differences between source and target datasets, thereby increasing the number of training samples available for target model training. 
Umer~\etal~\cite{umer2023imbalanced} use domain adaptation to deal with the problem of domain-specific data imbalances that may affect the generalization performance. 
Transfer learning methods usually rely on large-scale labeled auxiliary data to pre-train a model and have been widely used in various applications such as medical image classification~\cite{kim2022transfer}, brain abnormality identification~\cite{talo2019application}, and brain disease progression prediction~\cite{maqsood2019transfer}. 
Nanni~\etal\cite{nanni2020comparison} employ a transfer learning method for early diagnosis and prognosis of AD, by pre-training models on ImageNet~\cite{deng2009imagenet} that is further applied to MRI analysis. 
Bron~\etal~\cite{bron2021cross} validate that models trained for MRI-based AD identification can be transferred to predict future progression of mild cognitive impairment. 
Several recent studies~\cite{zhou2021models, zhou2021preservational} propose to employ a CT image reconstruction as an auxiliary task to pre-train models in an unsupervised manner. 
However, they generally require that source and target domains share identical label distributions for transfer learning. 
Inspired by these studies, we propose to use MRI reconstruction and brain tissue segmentation as auxiliary tasks for pre-training a model on large-scale MRIs and also transfer it to target domains (with different label distributions) for downstream prediction.

\section{Materials and Methodology}
\label{S3}
\subsection{Studied Subjects and Image Pre-Processing}
Three datasets with T1-weighted MRIs are included: (1) the public Alzheimer's Disease Neuroimaging Initiative (ADNI) dataset~\cite{jack2008alzheimer}, (2) a late-life depression (LLD) dataset~\cite{steffens2004methodology, steffens2017negative}, and (3) a diabetes mellitus (DM) dataset~\cite{tan2022convolutional}. 
In this work, ADNI is used as \emph{auxiliary source domain} for pre-training, while LLD and DM are used as \emph{target domains} for prediction.  

(1) \textbf{ADNI}. 
A total of 9,544 T1-weighted MRIs are downloaded from ADNI. 
These scans are collected at different time points from 2,370 subjects. 
These data are used as auxiliary source data to provide brain anatomy priors in this work. 

(2) \textbf{LLD}. 
A total of $309$ subjects from LLD are involved in this work.  
All participants of the LLD study are enrolled in two related studies: the Neurocognitive Outcomes of Depression in the Elderly study (NCODE)~\cite{steffens2004methodology} and the Neurobiology of Late-life Depression study (NBOLD)~\cite{steffens2017negative}. 
As shown in Table~\ref{Diagnosis_information}, these subjects are grouped into three categories: cognitively normal (CN), cognitively normal with depression (CND), and cognitive impairment or dementia with depression (CID). 
The demographic information is also shown in Table~\ref{Diagnosis_information}.
Note that category labels in LLD were determined based on \emph{5-year follow-up diagnosis}, while the MRIs were acquired at baseline time. 
In this work, this dataset is used as target data for MRI-based progression prediction of CI. 

(3) \textbf{DM}. 
The subjects in DM~\cite{tan2022convolutional} are grouped into three categories, including CN, cognitively normal with diabetes mellitus (DM), and mild cognitive impairment with diabetes mellitus (MCI), as shown in Table~\ref{Diagnosis_information}. 
A total of $139$ subjects are involved in this work, with detailed demographic information shown in Table~\ref{Diagnosis_information}. 
Similar to LLD, this DM dataset is also used as target data for MRI-based MCI identification. 

\if false
\textbf{MRI Acquisition}.
The T1-weighted MRI scans in NCODE were acquired using a 3D axial TURBOFLASH sequence with TR/TE=22/7 $ms$, flip angle=25°, a 100 Hz/pixel bandwidth, a 256×256 matrix, a 256 $mm$ diameter field-of-view, 160 slices with a 1 $mm$ slice thickness and Nex=1 (no signal averaging), yielding an image with 1 $mm$ cubic voxels. 
All sMRI scans in NBOLD were acquired using a Skyra 3T scanner (Siemens, Erlangen, Germany) with 32 surface coils located at Olin Neuropsychiatric Research Center (ONRC). 
Five high-resolution axial T1-weighted magnetization-prepared rapid gradient-echo (MPRAGE) images were acquired parallel with the anterior commissure-posterior commissure (AC-PC) line. 
The acquisition parameters were TR/TE=2,200/2.88 $ms$, flip angle=13°, matrix = 220×320×208, and voxel size 0.8×0.8×0.8 $mm$.
\fi

\textbf{Image Pre-Processing}. 
All structural MRIs from the three datasets are minimally preprocessed, including (1) bias field correction, (2) segmentation, (3) registration to standard MNI space with the size of $181\times217\times181$ and the spatial resolution of $1\times1\times1$~$mm^3$. 
The minimally-processed MRIs are then fed into the proposed framework for brain anatomy prior modeling and disease progression prediction. 

\subsection{Proposed Method}
\if false
While annotating MRIs is often challenging in practice, many MRIs exist in large-scale public datasets, even without any task-specific category labels. 
These MRIs have diagnosis information of specific diseases (\eg, AD stages in ADNI), which limits the application on other tasks (\eg, LLD and DM) without the related diagnosis.
However, even without task-specific category labels, the anatomical features (\eg, GM and WM volumes of the brain) that characterize brain anatomy can be utilized for further study~\cite{elsayed2011region, magnin2009support}. 
Such brain anatomy prior can be modeled via auxiliary tasks and employed to improve learning performance intuitively. 
\fi

While annotating MRIs is often challenging in practice, there are a large number of MRIs in existing large-scale datasets such as ADNI. 
Even without category labels, brain anatomical structure information derived from MRIs can be employed as prior knowledge to boost learning performance intuitively. 
Accordingly, we propose a brain
anatomy prior modeling (BAPM) framework for progression prediction of cognitive impairment, incorporated with brain anatomy prior provided by two auxiliary tasks (\ie, brain tissue segmentation and MRI reconstruction).  
As shown in Fig.~\ref{figFramework}, the BAPM consists of (1) a \emph{pretext model} with two auxiliary tasks for MRI reconstruction and brain tissue segmentation, respectively,  and (2) a \emph{downstream model} for prediction, both equipped with brain anatomy-guided encoders (shared parameters) for MRI feature learning. 
The BAPM can pre-train the encoder on large-scale auxiliary MRI data via two auxiliary tasks to model brain anatomy prior without diagnostic labels as supervision.
The encoder is then transferred to the downstream model, which is further fine-tuned on small-sized labeled target MRIs. 

\subsubsection{Pretext Model for Anatomy Prior Learning} 
To explicitly model brain anatomical structures in MRIs, we first design a pretext model (see Fig.~\ref{figFramework}), consisting of (1) \emph{a brain anatomy-guided encoder} for brain anatomy learning and (2) \emph{two decoders} for MRI reconstruction and brain tissue segmentation. 

The {brain anatomy-guided encoder} can take large-scale auxiliary 3D MRIs without category labels as input. 
It contains eight convolution blocks, with each block containing a convolution layer (kernel size: $3\times3\times3$), an instance normalization layer, and a parametric rectified linear unit (PReLU) activation. 
The channel numbers of the eight blocks are [64, 128, 256, 512, 512, 512, 1,024, 1,024], respectively. 
The first four blocks downsample the input with a stride of $2\times2\times2$.
A skip connection is applied to sum the input and output of every two of the last four blocks for residual learning. 
Finally, the encoder outputs two 512-dimensional feature maps. 

We also design two decoders (sharing the same architecture) to perform MRI reconstruction and tissue segmentation, respectively.
Each decoder takes the feature map generated by the encoder as input and outputs reconstructed MR images or segmentation maps of three types of brain tissues (\ie,  gray matter (GM), white matter (WM), and cerebrospinal fluid (CSF)), thus guiding the encoder to learn brain anatomy prior. 
Note that the MRI reconstruction task does not require any label information, while the brain tissue segmentation task relies on ground-truth segmentation maps. 

The decoder contains four deconvolution blocks, and each deconvolution block shares the same architecture as the convolution block in the encoder. 
The channel numbers of the deconvolution are [256, 128, 64, 1] for reconstruction and [256, 128, 64, 4] for segmentation, respectively.
The output of each decoder is then fed into a SoftMax layer for image reconstruction or tissue segmentation. 
The reconstruction task can be used to train the pretext model in a \emph{totally unsupervised manner}, and we use the $l_1$-norm as the loss function: 
\begin{equation}
\small
 L_{rec} = \frac{1}{D}\sum\nolimits_{i=1}^{D}||x_i-\hat{x}_i||_1,
\end{equation}
where $x_i$ and $\hat{x}_i$ denote the $i$-th voxel in the input MRI and the reconstructed image, respectively, and $D$ is the number of voxels in an MRI. 
In the segmentation task, the corresponding decoder produces four probability maps indicating the probability of a voxel belonging to a specific tissue (\ie, background, WM, GM, and CSF), with a Dice coefficient-based loss function defined as: 
\begin{equation}
\small
 L_{seg} = -\frac{2\sum_{i=1}^{D}y_i\hat{y}_i}{\sum_{i=1}^{D}y_i^2+\sum_{i=1}^{D}\hat{y}_i^2},
\end{equation}
where $y_i$ is the segmentation output, $\hat{y}_i$ is the ground truth.

The loss function of the pretext model is formulated as: 
\begin{equation}
\small
L = L_{rec} + L_{seg}. 
\end{equation}
Particularly, the segmentation task needs ground-truth segmentation maps of brain tissues for training data, and many established tools such as FSL
, FreeSurfer
, SPM 
and iBEAT~\cite{wang2023ibeat} can be used. 
In this work, we use iBEAT with careful manual verification to generate such ground-truth segmentation maps, aiming to provide more accurate segmentation. 
More discussions can be found in Section~\ref{S_segResult}.

To increase diversity of input data, each input source MRI is further augmented through random affine transformation, random blur, random noise, random bias field, and random motion artifact to simulate 
image quality variance, magnetic field inhomogeneity, and
motion artifacts, when training the pretext model. 
Discussions on the influence of these two auxiliary tasks and data augmentation can be found in Section~\ref{S5}.

\subsubsection{Downstream Model for Target Prediction}
As shown in the bottom panel of Fig.~\ref{figFramework}, the downstream model takes target MRIs as input and outputs predicted category labels. 
It consists of (1) a \emph{brain anatomy-guided encoder} and (2) a \emph{predictor} for forecasting. 
To address the small-sample-size issue, we propose to first pre-train the encoder through the above-mentioned pretext model on large-scale public MRIs, and then transfer it to the downstream model. 

With small-sized labeled target MRIs as training data, we further fine-tune the downstream model in a supervised learning manner (with the encoder frozen). 
Here, the predictor has two branches to learn the features guided by the two auxiliary tasks separately, each having two convolution blocks (kernel size: $3\times3\times3$, stride:  $2\times2\times2$, channel: 256) with a skip connection.
Specifically, the two 512-dimensional features generated by the pre-trained encoder are first fed into two parallel branches for feature abstraction, followed by a sum operation and a fully-connected layer for feature fusion and prediction. 
It's worth noting that the architecture of the predictor can be flexibly adjusted according to the requirements of different downstream tasks (\eg, using more complex architectures for problems with a larger number of labeled target samples). 
For the downstream task on target data, we use a cross-entropy loss for optimization.

It's worth noting that, for source MRIs used in the pretext model, the MRI reconstruction task does not require any label information, while the brain tissue segmentation task relies on ground-truth segmentation maps. 
In the downstream model, we require some category labels (\eg, CI) of target MRIs for fine-tuning. 
That is, auxiliary source MRIs and target MRIs used in the proposed BAPM can have totally different label distributions, thus significantly improving its practical utility.

\subsubsection{Implementation}
The proposed BAPM is trained using a two-step optimization strategy. 
(1) We first train the pretext model on 9,344 MRIs from ADNI, with ground-truth segmentation as supervision.  
The Adam optimizer~\cite{kingma2014adam} is used, with a learning rate of ${10}^{-4}$, batch size of 4, and training epoch of 30. 
(2) We then transfer the parameters of the encoder learned in pretext model to the downstream model and fine-tune the predictor on target data (batch size: 2, start learning rate: ${10}^{-4}$, epoch: 90). 
The learning rate of fine-tuning decays by 0.1 every 30 epochs.  
The BAPM is implemented on PyTorch 
with NVIDIA TITAN Xp (memory: 12GB). 

\section{Experiments}
\label{S4}

\subsection{Experimental Setup} 

Two types of binary classification tasks are performed: 
(1) CI recognition in the LLD study (\ie, CID vs. CN and CID vs. CND classification), and 
(2) MCI detection in the DM study (\ie, MCI vs. CN and  MCI vs. DM classification). 
The classification performance is evaluated through five evaluation metrics: area under the ROC curve (AUC), accuracy (ACC), sensitivity (SEN), specificity (SPE), and F1-Score (F1s). 
Since we only have limited and unbalanced samples in the target datasets for fine-tuning, we randomly select 80\% subjects from each category for training and the remaining 20\% for testing.   
The experiments run five times independently to avoid any bias introduced by the random splitting of training/test set, and the mean and standard deviation of five metrics are reported.
For two target datasets (\ie, LLD and DM), the training data is duplicated and augmented using a random affine transform.

\subsection{Competing Methods}
We compare our BAPM with the most popular machine learning methods for downstream prediction tasks, including (1) support vector machine (\textbf{SVM}) with a radial basis function kernel~\cite{pisner2020support}, and (2) extreme gradient boosting (\textbf{XGB})~\cite{chen2016xgboost} (estimator: 300, tree depth: 4, learning rate: 0.2). 
These two classifiers take handcrafted MRI features as inputs, including average image intensities of gray matter (GM) and white matter (WM) within pre-defined 166 regions-of-interest (ROIs) in AAL3~\cite{rolls2020automated} (denoted as \textbf{SVM/XGB-GM} and \textbf{SVM/XGB-WM}, respectively). 
Specifically, in these four methods, we further process the minimally-processed MRIs to extract ROI features, including deformable registration to AAL3, 
ROI partition of the registered MRIs based on the AAL3 template, and ROI-based GM and WM feature extraction.

We also compare the BAPM with the following six state-of-the-art deep learning methods, with details introduced below. 

(1) \textbf{MobileNet}~\cite{howard2017mobilenets}: This is a lightweight 3D convolutional neural network (CNN) model, with 28 convolutional layers. 

(2) \textbf{EfficientNet}~\cite{tan2019efficientnet}: It is a CNN scaling method that uniformly scales up the model's all dimensions using a compound coefficient, with the EfficientNet-B0 model used in this work. 

(3) \textbf{ResNet}~\cite{he2016deep}: ResNet is a popular CNN-based model that stacks residual blocks on top of each other to form a network. 
ResNet has many variants with similar architecture but different numbers of layers, from 10 to 200 layers. 
We compare the 3D version of ResNet18, ResNet34, and ResNet50, considering our input size and GPU memory capacity. 

(4) \textbf{SEResNet}~\cite{hu2018squeeze}: SEResNet is an improved ResNet model by adding squeeze and excitation blocks to ResNet, and the SEResNet50 is used for comparison. 

(5) \textbf{Med3D}~\cite{chen2019med3d}: The Med3D is pre-trained on segmentation of multiple organs based on 1,474 3D MRI and CT images, and fine-tuned on our target MRI. 
We download the pre-trained three models (\ie, {Med3D18, Med3D34, and Med3D50}) from GitHub\footnote{https://github.com/Tencent/MedicalNet}, and fine-tune them on target data for prediction. 

(6) \textbf{DeepTransfer}~\cite{bron2021cross}: DeepTransfer is a transfer learning method that uses AD vs. CN classification as the auxiliary task and transfers the pre-trained model for downstream prediction. 
Specifically, it is pre-trained on ADNI to identify AD patients from CNs (with 60 AD and 710 CN) in a supervised learning manner, where 10\% of subjects are left out for validation. 
After pre-training, this model is further fine-tuned on target data for prediction. 
For a fair comparison, it shares the same architecture as the downstream model in our BAPM.

Note that the five methods (\ie, Med3D18, Med3D34, Med3D50, DeepTransfer, and our BAPM) use different pre-training strategies, while the remaining ones only utilize target data for model training. 
For each downstream/target prediction task in this work, all the ten deep learning methods take whole 3D MRIs as input and share the same training/fine-tuning and data augmentation strategies as that used in the downstream model of our BAPM. 
In the experiments, we typically use the default setting of all competing methods and make a concerted effort to ensure that the network architecture and hyperparameters are comparable to the proposed BAPM.

\begin{table*}[!t]
\setlength{\belowdisplayskip}{0pt}
\setlength{\abovedisplayskip}{0pt}
\setlength{\abovecaptionskip}{0pt}
\setlength{\belowcaptionskip}{0pt}
\footnotesize
\centering
\renewcommand{\arraystretch}{1}
\caption{Performance of fourteen methods in two CI recognition tasks (\ie, CID vs. CND and CID vs. CN classification) on LLD, with `*' denoting the results of BAPM and a competing method are statistically significantly different ($p < 0.05$).}
\label{comparison_lld}
\setlength{\tabcolsep}{2pt}
\begin{tabular}{l|ccccc|ccccc}
\toprule
\multirow{2}{*}{Method} &\multicolumn{5}{c|}{CID vs. CND Classification} &\multicolumn{5}{c}{CID vs. CN Classification}\\
\cmidrule(lr){2-6} \cmidrule(lr){7-11} 
&AUC (\%) &ACC (\%) &SEN (\%) &SPE (\%)  &F1s (\%)  &AUC (\%) &ACC (\%) &SEN (\%)  &SPE (\%)  &F1s(\%) \\
\midrule
SVM-GM    &47.00±4.14 &52.00±5.42* &56.00±15.57 &48.00±5.70 &53.12±9.30 
&51.78±4.10 &49.50±3.71* &49.00±5.48 &50.00±11.73 &49.17±2.12             \\
SVM-WM    &47.62±6.79 &49.00±3.79* &50.00±16.58 &48.00±12.55 &48.33±10.62 
&51.78±1.89 &50.50±2.09* &49.00±9.62 &52.00±10.95 &49.36±5.10             \\
XGB-GM    &50.30±15.48 &49.50±15.35* &53.00±17.54 &46.00±14.75 &50.95±16.04 
&47.10±6.76 &49.00±2.85* &51.00±12.94 &47.00±14.40 &49.31±7.42            \\
XGB-WM    &51.55±6.22 &49.00±8.40* &50.00±9.35 &48.00±9.08 &49.44±8.64
&45.90±8.43 &46.00±9.78* &46.00±15.97 &46.00±4.18 &45.34±13.30             \\

\hline
MobileNet   &56.80±11.22 &55.00±7.71* &45.00±14.58 &65.00±14.14 &49.35±9.98             &50.90±13.82 &53.00±10.37 &57.00±17.54 &49.00±7.42 &54.14±12.35 \\
            
EfficientNet 
            &64.75±4.13 &58.50±4.54 &35.00±5.00 &82.00±9.75 &45.71±4.77 
            &56.95±9.96 &51.00±4.87 &65.00±5.00 &37.00±11.51 &57.03±2.91\\ 
            
SEResNet   
            &59.60±9.15 &55.50±8.37* &35.00±16.96 &76.00±4.18 &42.75±14.59 
            &58.95±6.97 &57.00±6.47 &55.00±8.66 &59.00±6.52 &56.00±7.13\\
            
ResNet18    &64.35±4.77 &58.50±6.75* &41.00±7.42 &76.00±10.84 &49.62±7.86 
            &62.15±4.07 &57.00±4.47* &53.00±17.89 &61.00±13.42 &54.08±9.94\\
            
ResNet34   &63.50±20.98 &58.50±11.54* &35.00±14.14 &82.00±14.40 &45.17±17.71 
            &59.90±6.53 &55.50±4.11 &54.00±7.42 &57.00±10.95 &54.70±3.95\\            
            
ResNet50   &65.10±13.69 &60.00±3.06* &38.00±9.75 &82.00±7.58 &48.13±7.99 
            &59.40±15.12 &59.00±9.94* &53.00±11.51 &\textbf{65.00±13.23} &56.21±11.08 \\

Med3D18    &63.25±8.68 &57.50±4.68* &39.00±6.52 &76.00±13.42 &47.69±3.94 
            &61.45±6.14 &57.00±5.12 &58.00±16.43 &56.00±10.84 &56.65±8.71\\
            
Med3D34  &65.70±5.36 &60.50±8.91 &47.00±13.96 &74.00±9.62 &53.82±11.84 
            &60.55±8.37 &58.00±4.11 &56.00±14.32 &60.00±11.73 &56.44±7.72\\
           
Med3D50   &56.10±8.77 &54.00±2.24 &39.00±5.48 &69.00±9.62 &45.70±2.70 
            &58.80±6.01 &56.50±2.85* &53.00±7.58 &60.00±10.00 &54.75±3.62\\ 

DeepTransfer   &66.40±4.66 &61.00±4.54* &46.00±15.57 &76.00±10.84 &52.74±12.82 
            &64.65±5.17 &61.00±2.85 &65.00±11.73 &57.00±11.51 &62.08±5.60 \\%
\hline
BAPM~(Ours)            
            &\textbf{75.10±4.65} &\textbf{69.00±5.18} &\textbf{53.00±12.04} &\textbf{85.00±6.12} &\textbf{62.52±8.69} 
            &\textbf{64.65±3.28} &\textbf{63.50±2.85} &\textbf{67.00±9.08} &60.00±10.61 &\textbf{64.54±3.74} \\

\bottomrule
\end{tabular}
\end{table*}

\begin{table*}[!t]
\setlength{\belowdisplayskip}{0pt}
\setlength{\abovedisplayskip}{0pt}
\setlength{\abovecaptionskip}{0pt}
\setlength{\belowcaptionskip}{0pt}
\footnotesize
\centering
\renewcommand{\arraystretch}{1}
\caption{Performance of fourteen methods in two MCI recognition tasks (\ie, MCI vs. CN and MCI vs. DM classification) on DM, with `*' denoting the results of BAPM and a competing method are statistically significantly different ($p < 0.05$).}
\label{comparison_dm}
\setlength{\tabcolsep}{2pt}
\begin{tabular}{l|ccccc|ccccc}
\toprule
\multirow{2}{*}{Method} &\multicolumn{5}{c|}{MCI vs. CN Classification} &\multicolumn{5}{c}{MCI vs. DM Classification}\\
\cmidrule(lr){2-6} \cmidrule(lr){7-11} 
&AUC (\%) &ACC (\%) &SEN (\%) &SPE (\%)  &F1s (\%)  &AUC (\%) &ACC (\%) &SEN (\%)  &SPE (\%)  &F1s(\%) \\
\hline
SVM-GM    &56.70±8.60 &53.50±6.52* &65.00±11.18 &42.00±9.75 &58.05±6.96 
            &57.72±6.16 &56.50±5.48* &58.00±6.71 &55.00±10.00 &57.10±4.90 \\
SVM-WM    &51.08±8.35 &47.00±4.81* &50.00±11.18 &44.00±10.84 &48.13±6.99 
            &53.78±6.05 &51.50±4.54* &53.00±7.58 &50.00±12.75 &52.10±3.77 \\
XGB-GM   &54.80±3.62 &56.50±8.22* &56.00±10.84 &57.00±7.58 &56.09±9.28 
            &51.80±9.84 &50.00±7.91* &53.00±13.96 &47.00±9.75 &50.97±9.90 \\
XGB-WM    &59.12±7.41 &55.00±3.54* &57.00±13.51 &53.00±12.55 &55.20±7.57 
            &50.50±6.99 &48.00±6.94* &51.00±9.62 &45.00±9.35 &49.33±7.55 \\
\hline
MobileNet   
            &54.05±8.77 &56.50±8.22* &57.00±10.37 &56.00±9.62 &56.58±8.89 
            &49.05±5.85 &50.50±7.58* &54.00±8.94 &47.00±8.37 &52.10±7.66 \\
EfficientNet &59.60±8.55 &56.50±6.02* &59.00±12.94 &54.00±8.22 &57.11±8.23 
            &49.70±5.23 &48.50±6.75* &51.00±8.94 &46.00±8.94 &49.63±7.24 \\
            &58.75±6.18 &55.50±5.97* &56.00±5.48 &55.00±7.91 &55.75±5.41 
            &57.85±9.61 &56.50±6.98 &52.00±13.51 &61.00±6.52 &53.86±10.53 \\
ResNet18 &59.10±6.77 &55.00±5.59 &52.00±13.04 &58.00±10.37 &53.02±8.70 
            &63.70±6.33 &56.50±4.54* &54.00±6.52 &59.00±7.42 &55.31±5.08 \\
ResNet34   &62.55±8.70 &60.00±4.68 &56.00±11.40 &64.00±7.42 &57.91±7.30 
            &58.20±10.02 &58.00±10.81* &56.00±7.42 &60.00±15.00 &57.43±9.77 \\
ResNet50   &60.35±6.19 &57.00±5.42* &58.00±12.04 &56.00±9.62 &57.00±7.69 
            &57.65±6.21 &56.00±2.85* &59.00±2.24 &53.00±7.58 &57.31±1.02 \\
Med3D18 &61.85±7.09 &58.50±5.18* &66.00±5.48 &51.00±8.22 &61.40±4.51
            &55.30±9.88 &52.00±5.12* &48.00±6.71 &56.00±8.22 &49.93±5.28 \\
Med3D34 &60.95±7.70 &55.50±6.22 &50.00±10.00 &61.00±6.52 &52.62±8.51 
            &60.10±13.54 &59.00±9.12* &61.00±11.40 &57.00±10.37 &59.66±9.42 \\
Med3D50      &55.85±6.47 &56.00±6.75* &57.00±13.04 &55.00±6.12 &55.98±9.27 
            &54.55±4.59 &53.00±4.11* &57.00±10.37 &49.00±8.22 &54.47±6.53 \\
DeepTransfer   &60.70±12.38 &56.00±10.09 &50.00±8.66 &62.00±13.51 &53.31±9.45 
&56.95±4.53 &54.50±6.94 &49.00±6.52 &60.00±11.73 &51.89±6.29 \\

\hline
BAPM~(Ours)   &\textbf{67.75±8.55} &\textbf{66.50±9.45} &\textbf{67.00±14.40} &\textbf{66.00±7.42} &\textbf{66.25±10.70} 
            &\textbf{65.55±5.62} &\textbf{64.00±8.40} &\textbf{62.00±7.58} &\textbf{66.00±11.94} &\textbf{63.35±7.77} \\
\bottomrule
\end{tabular}
\end{table*}

\subsection{Results of CI Detection on LLD Study} 
We first report the results achieved by the proposed BAPM and the competing methods in two downstream prediction tasks (\ie, CID vs. CND classification, CID vs. CN classification) in Table~\ref{comparison_lld}. 
Note that there are much fewer CID subjects in this study, compared to CND and CN categories. 
From Table~\ref{comparison_lld}, we have the following observations.

\emph{First}, our BAPM generally outperforms the competing methods in most cases. 
Significantly, the BAPM achieves the highest SEN value (\ie, $67.00\%$) in CID vs. CN classification, suggesting that our method is effective in identifying depressed subjects who progress to CI within five years from healthy subjects. 
\emph{Besides}, compared with methods with model pre-training (\ie, Med3D18, Med3D34, and Med3D50), BAPM and DeepTransfer produce the overall better results in two tasks. 
This may be due to the fact that BAPM and DeepTransfer are pre-trained on brain MRI that share a similar distribution as target data, but Med3D18, Med3D34 and Med3D50 are pre-trained on MRI and CT scans of multiple organs. 
It suggests that the distribution gap between source and target data can greatly affect model's transferable ability. 
\emph{Additionally}, BAPM is superior to DeepTransfer in most cases, possibly due to that brain anatomy prior modeled by two auxiliary tasks helps boost learning performance. 
\emph{Furthermore}, the impact of data imbalance on deep learning methods is obviously observed in terms of the F1s results, and the proposed BAPM yields the highest F1 scores in both tasks. 

\begin{figure*}[!t]
\setlength{\belowdisplayskip}{0pt}
\setlength{\abovedisplayskip}{0pt}
\setlength{\abovecaptionskip}{0pt}
\setlength{\belowcaptionskip}{0pt}
\centering
\includegraphics[width=1\textwidth]{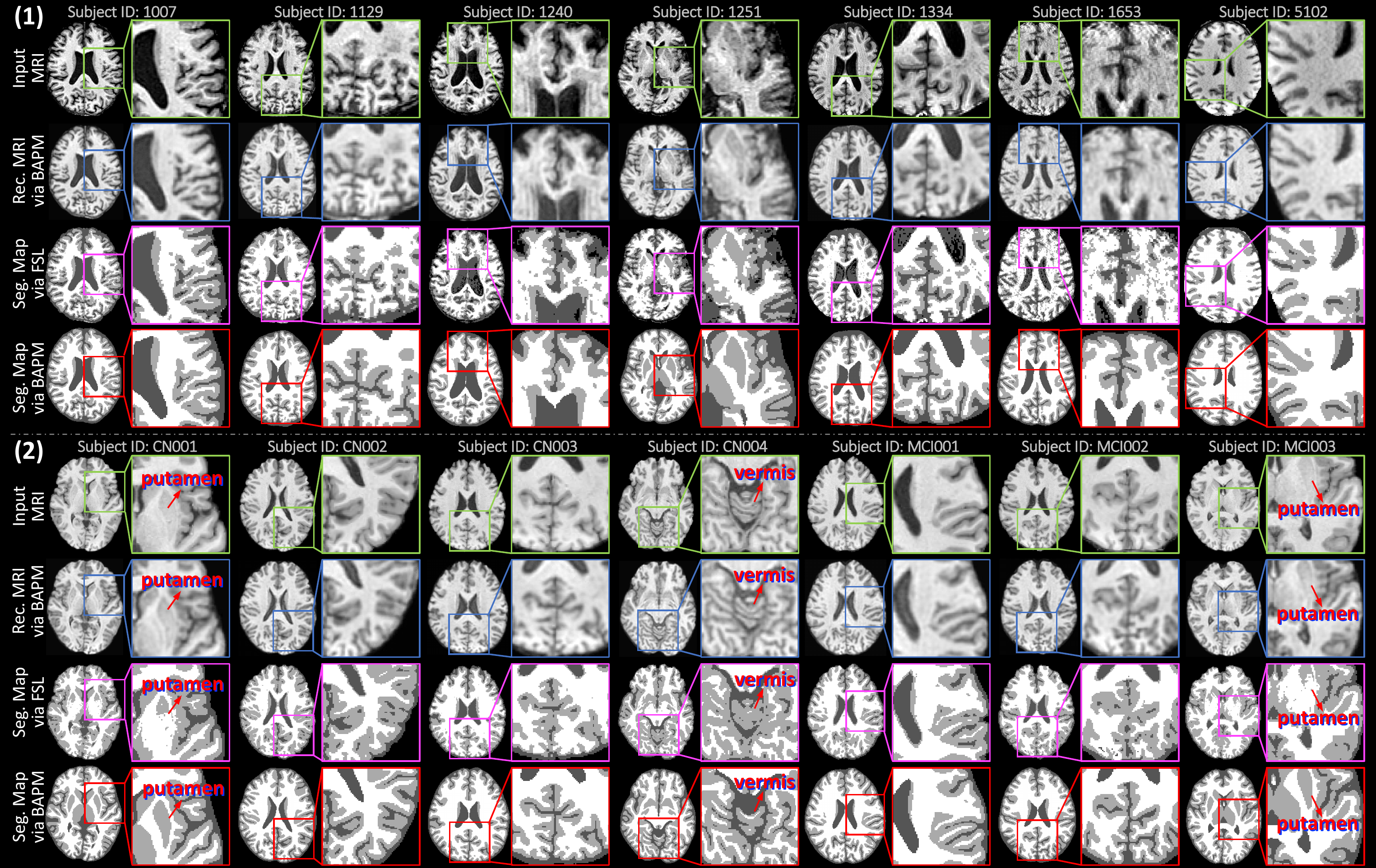}
\caption{MRI reconstructions (rec.) and tissue segmentation (seg.) maps of white matter, gray matter and cerebrospinal fluid 
produced our BAPM and FSL on (1) LLD study and (2) DM study.}
\label{figSeg}
\end{figure*}
\subsection{Results of MCI identification on DM Study} 
The results achieved by the BAPM and the competing methods in two MCI detection tasks (\ie, MCI vs. CN classification, MCI vs. DM classification) on the DM study are reported in Table~\ref{comparison_dm}. 
We have more balanced data in each category in the DM study for model fine-tuning 
than LLD. 
From Table~\ref{comparison_dm}, we have a similar observation to those in Table~\ref{comparison_lld}, that is, the proposed  BAPM produces the overall best performance in most cases. 
\emph{On the other hand}, the deep learning methods with model pre-training (\eg, Med3D34, DeepTransfer, and our BAPM) usually produce better results than those without pre-training. 
For instance, our BAPM achieves a significant improvement of $6.50\%$ in terms of ACC, compared with that of ResNet34 (ACC=$60.00\%$). 
This further validates that exploring auxiliary source data for model pre-training can help boost the performance of downstream prediction tasks. 
\if false
we can see that BAPM is superior to DeepTransfer in most cases, possibly due to that brain anatomy prior modeled by two auxiliary tasks helps boost learning performance.
\emph{First}, our BAPM outperforms all competing methods in most cases. 
The BAPM outperforms the second-best methods on AUC, ACC, and F1s in both identification tasks. 
These results show the effectiveness of BAPM for MCI identification in the DM study. 
\fi 
\emph{Besides}, with more balanced training data for fine-tuning, BAPM produces higher F1s results in the DM study than in the LLD study. 
This implies that data imbalance may be a significant issue affecting the performance of deep learning models when the number of training samples is limited.

\subsection{Reconstruction and Segmentation Results} 
\label{S_segResult}
In the proposed BAPM, the pre-trained pretext model can also be used for MRI reconstruction and brain tissue segmentation in downstream studies. 
So, we further \emph{qualitatively} and \emph{quantitatively} evaluate the performance of the pretext model via two groups of experiments. 
We first visualize the results of our pretext model on target MRIs from LLD and DM studies in MRI reconstruction and tissue segmentation in Fig.~\ref{figSeg}.
The brain segmentation maps generated by FSL are also visualized for comparison. 
Note that T1-weighted MRIs in the LLD study are collected from two sites and have more inconsistent image quality when compared to those from DM. 

From Fig.~\ref{figSeg}, we have several exciting observations. 
\emph{First}, our reconstructed MRIs have more consistent image quality than the original ones. 
For instance, our methods help remove some noise (see ID: 1251 from LLD) and motion artifacts (see ID: 1653 from LLD) in the reconstructed images. 
This may imply that using MRI reconstruction as an auxiliary task can guide the encoder to extract more general features that are resistant to noise and artifacts. 
\emph{Second}, the segmentation maps generated by our pretext model are generally better than those of FSL in most cases, especially for those \emph{cortical surface areas} in the two studies. 
For instance, the WM region in segmentation maps generated by BAPM is much cleaner than that of FSL, indicating that our model is not sensitive to noise in MRI. 
Even for the LLD study with significant inter-site data heterogeneity, 
 the boundary of WM and GM produced by BAPM is more continuous and smoother, which is in line with brain anatomical structures. 
\emph{Besides}, for MRIs (IDs: 1240, 1334, and 1653 from LLD) with severe motion artifacts, our model can produce high-quality segmentation maps that are even comparable to those of MRIs without motion artifacts.
This demonstrates that our model is robust to noise and motion artifacts in both reconstruction and segmentation tasks. 
The underlying reason could be that the pretext model is trained on large-scale MRIs and thus has good generalization ability when applied to MRIs with different image quality. 
\emph{Additionally}, our BAPM can achieve better segmentation results in many fine-grained brain regions than FSL, such as the \emph{putamen region} (see CN001 and MCI003 from DM) and the \emph{vermis region} (see CN004 from DM). 
These results demonstrate the excellent adaptability of the pretext model when applied to downstream tasks for image reconstruction and tissue segmentation. 

\begin{table}[!t]
\setlength{\belowdisplayskip}{0pt}
\setlength{\abovedisplayskip}{0pt}
\setlength{\abovecaptionskip}{0pt}
\setlength{\belowcaptionskip}{0pt}
\scriptsize
\centering
\caption{The MRI reconstruction and tissue segmentation performance of BAPM and FSL.}
\label{tab_rec_seg}
\setlength{\tabcolsep}{1pt}
\begin{tabular}{l|ccc|ccc}
\toprule
\multirow{2}{*}{Method} &\multicolumn{3}{c|}{MRI Reconstruction} &\multicolumn{3}{c}{Brain Tissue Segmentation}\\
\cmidrule(lr){2-4} \cmidrule(lr){5-7} 
&MAE &NMI (\%) &SSIM (\%) & Dice (\%) &ASD (mm) &HD (mm) \\
\midrule
FSL &-  &- &- &82.30±1.57  &0.81±0.14  &12.29±2.40 \\ 
BAPM  &0.0155±0.0017 &99.05±8.17 &97.24±0.40
&90.94±0.48 &0.37±0.03 &9.63±2.82\\ 
\bottomrule
\end{tabular}
\end{table}
In the second experiment for \emph{quantitative evaluation}, we apply the pretext model to 200 MRIs from ADNI for MRI reconstruction and tissue segmentation, while these test MRIs are independent of training data. 
Similar to the training MRIs, the ground-truth tissue segmentation maps of these test MRIs are generated using iBEAT with manual verification. 
The performance of MRI reconstruction is evaluated using mean absolute error (MAE), normalized mutual information (NMI), and structural similarity index (SSIM).
We also compare our method with FSL~\cite{jenkinson2012fsl} for tissue segmentation using Dice, average surface distance (ASD), and Hausdorff distance (HD). 
The experimental results are reported in Table~\ref{tab_rec_seg}. 
The results show that the pretext model trained on large-scale MRIs can reconstruct the input MRIs with high similarity (\ie, NMI of 99.05\% and SSIM of 97.24\%). 
For tissue segmentation, our method consistently outperforms FSL in terms of three metrics. 
Their results are consistent with those in Fig.~\ref{figSeg}.

\section{Discussion}
\label{S5}
In this section, we analyze and discuss the influences of several important components of the proposed BAPM method. 
We also discuss the limitations of the current work and present several future research directions. 

\begin{figure}[!t]
\setlength{\belowdisplayskip}{0pt}
\setlength{\abovedisplayskip}{0pt}
\setlength{\abovecaptionskip}{0pt}
\setlength{\belowcaptionskip}{0pt}
\centering
\includegraphics[width=0.48\textwidth]{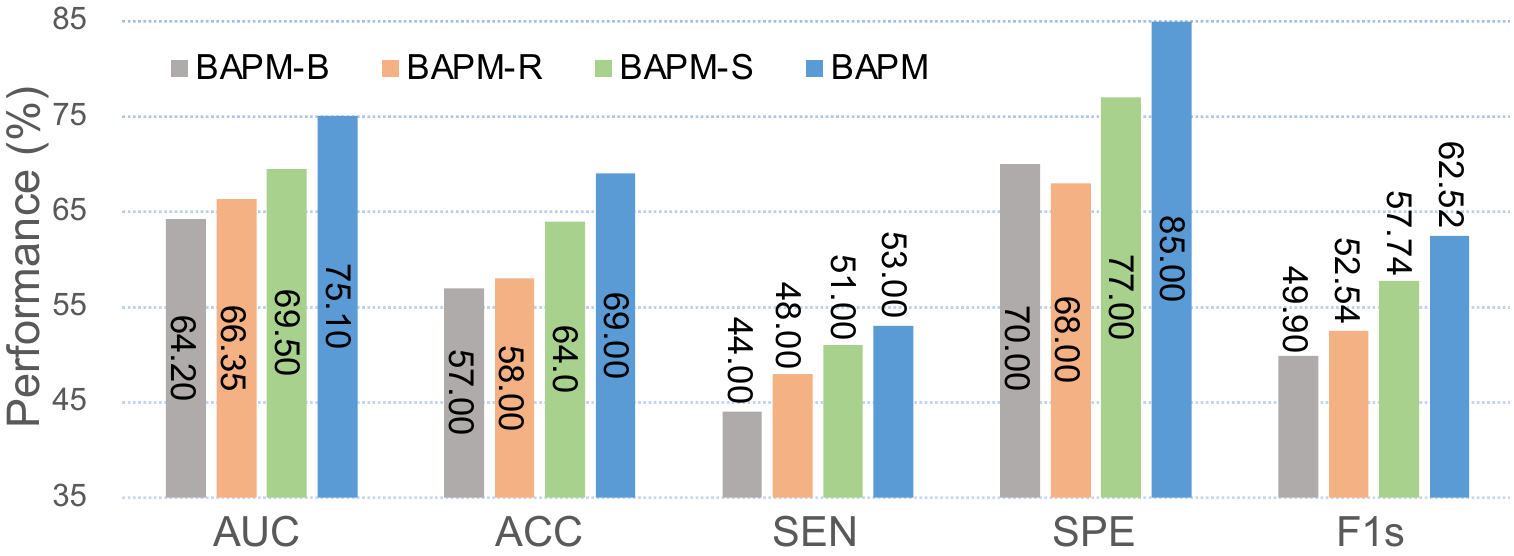}
\caption{Results achieved by BAPM with different auxiliary tasks in CID vs. CND classification on LLD. 
BAPM-B denotes the downstream model trained from scratch. 
BAPM-R and BAPM-S denote the pretext model trained with MRI reconstruction and brain tissue segmentation separately. 
} 
\label{fig-ablation}
\end{figure}
\subsection{Effectiveness of Anatomy Prior Modeling Strategies}
To validate the effectiveness of the proposed anatomy prior modeling strategy, 
we compare BAPM with its two variants (called \textbf{BAPM-R}, and \textbf{BAPM-S}) that model brain anatomy prior through different auxiliary tasks. 
Specifically, the BAPM-R trains the pretext model through an \emph{MRI reconstruction task} in a fully unsupervised learning manner. 
The BAPM-S trains the pretext model through a \emph{brain tissue segmentation task} using tissue segmentation maps as supervision. 
The results of these methods in CID vs. CND classification on LLD are reported in Fig.~\ref{fig-ablation}.  
We also report the results using the model (called \textbf{BAPM-B}) \emph{trained from scratch} on target data. 
As shown in Fig.~\ref{fig-ablation}, BAPM consistently performs better than its variants in terms of all five metrics. 
This implies that the proposed two auxiliary tasks help improve the discriminative ability of MRI features to boost prediction performance. 
Besides, BAPM-S is superior to BAPM-R in most cases, implying that brain anatomy prior derived from tissue  
segmentation is more useful in improving the generalizability of the pre-trained encoder when compared with MRI reconstruction.

\begin{figure}[!t]
\setlength{\belowdisplayskip}{0pt}
\setlength{\abovedisplayskip}{0pt}
\setlength{\abovecaptionskip}{0pt}
\setlength{\belowcaptionskip}{0pt}
\centering
\includegraphics[width=0.48\textwidth]{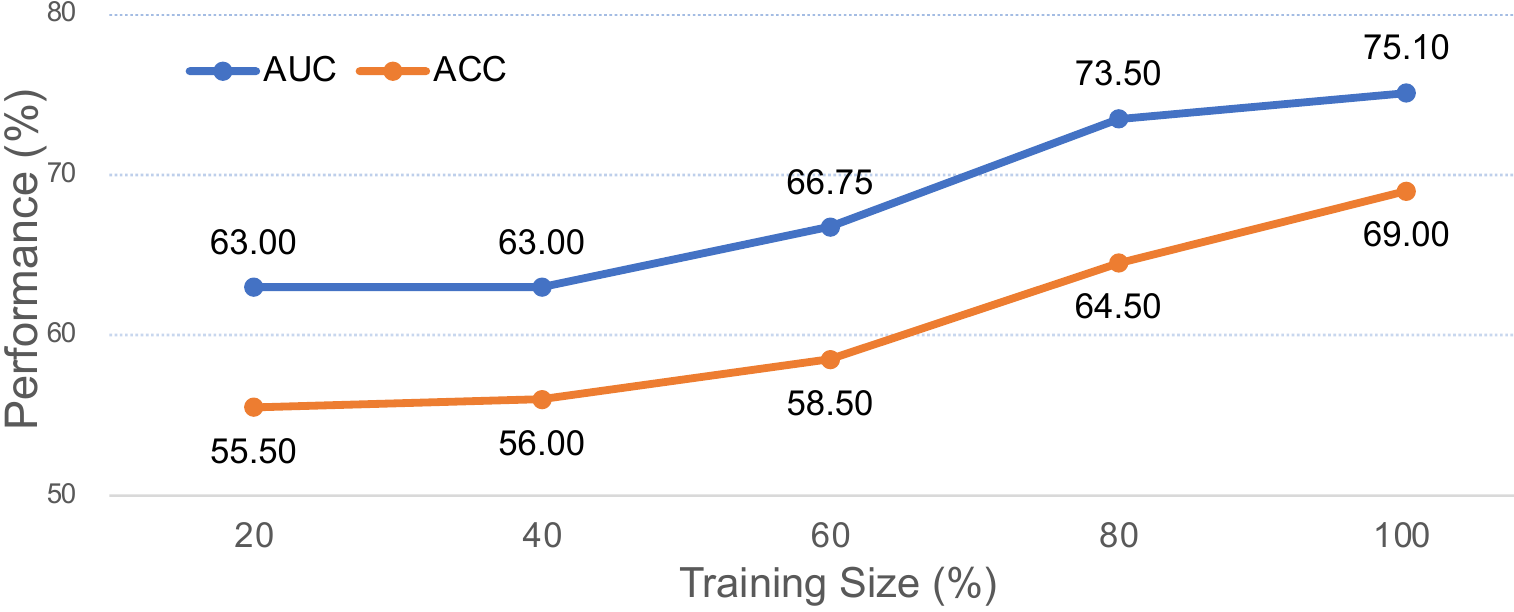}
\caption{Performance of BAPM with different numbers of source data for pre-training in CID vs. CND classification. The training size denotes the percentages of the total training data used for pretext model training.} 
\label{fig-trainsize}
\end{figure}
\subsection{Influence of Source Data Size on Pretext model}
We also study the influence of source data size on BAPM in CID vs. CND classification on LLD, with results shown in Fig.~\ref{fig-trainsize}. 
Specifically, we first randomly select subsets (\ie, $[20\%, 40\%, \cdots, 100\%]$ of 9,344 MRI scans) from ADNI to train five pretext models, and then transfer these encoders to downstream models for prediction. 
It can be observed from Fig.~\ref{fig-trainsize} that the overall performance of our BAPM rises with the increase of source data, and it produces the best results when using all source data for pretext model training. 
When we use 80\% of all source MRIs, our BAPM can produce reasonable results. 
This suggests that using more data for pretext model training helps promote learning performance, and the benefits brought by source data increase are not that obvious when source samples reach a certain scale.

\begin{figure}[!t]
\setlength{\belowdisplayskip}{0pt}
\setlength{\abovedisplayskip}{0pt}
\setlength{\abovecaptionskip}{0pt}
\setlength{\belowcaptionskip}{0pt}
\centering
\includegraphics[width=0.48\textwidth]{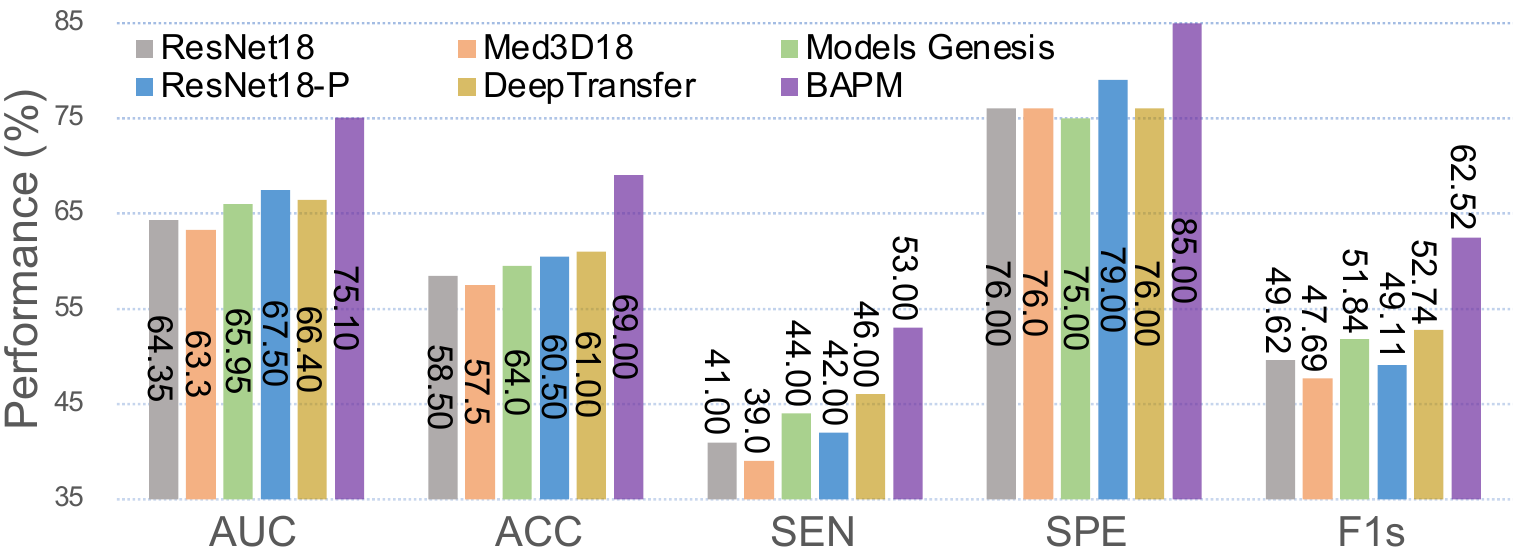}
\caption{Results of different methods with different pre-training strategies in CID vs. CND classification.} 
\label{fig_pretraining}
\end{figure}

\subsection{Impact of Model Pre-Training Strategy}
To study the impact of pre-training strategies, we compare our BAPM with four state-of-the-art (SOTA) methods, including Med3D18, ResNet18-P, DeepTransfer and Model Genesis~\cite{zhou2021models}. 
The Med3D18 is pre-trained on segmentation of multiple organs with 1,474 MRI or CT scans, and fine-tuned on our target MRI. 
Similar to DeepTransfer, ResNet18-P is first pre-trained on ADNI for AD vs. CN classification and then fine-tuned on the target LLD study (with the first three residual blocks frozen)~\cite{bron2021cross}. 
Models Genesis is pre-trained through an image reconstruction task on 623 Chest CT scans\footnote{https://github.com/MrGiovanni/ModelsGenesis/tree/master/pytorch}, which is fine-tuned on the target MRI data. 
That is, both Med3D18 and Models Genesis are only used to initialize the network parameters of the downstream model. 
The results of BAPM and four methods in the task of CID vs. CND classification on LLD are reported in Fig.~\ref{fig_pretraining}, where RestNet18 is used as a baseline without any pre-training.

Figure~\ref{fig_pretraining} suggests that even pre-trained on relatively larger-scale data, the two methods (\ie, Med3D18 and Models Genesis) still produce the overall poor performance than the three methods pre-trained on brain MRIs (\ie, RestNet18-P, DeepTransfer, and BAPM). 
This could be due to the fact that there are large data distribution differences between source and target images used in Med3D18 and DeepTransfer. 
Additionally, our BAPM consistently performs better than RestNet18-P and DeepTransfer. 
This implies that explicitly incorporating brain anatomy prior to model pre-training (as we do in BAPM) is effective to learn more general encoders for performance improvement in downstream tasks. 
This further validates the rationality of our motivation in this work. 

\subsection{Impact of Data Augmentation on Segmentation}
\begin{figure}[!t]
\setlength{\belowdisplayskip}{0pt}
\setlength{\abovedisplayskip}{0pt}
\setlength{\abovecaptionskip}{1pt}
\setlength{\belowcaptionskip}{0pt}
\centering
\includegraphics[width=0.48\textwidth]{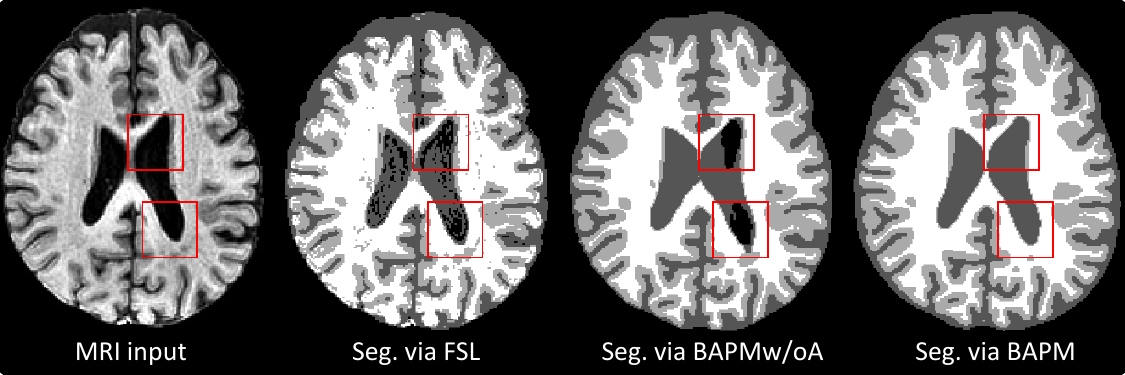}
\caption{Segmentation (Seg.) results of three methods. 
}
\label{figDataAug_on_Seg}
\end{figure}

Inspired by previous studies~\cite{zhang2023input,iglesias2023synthsr}, we augment the source MRIs using random affine transformation, random blur, random noise, random bias field, and random motion artifact. 
We now study the effect of such data augmentation strategy by comparing our BAPM with its variant (called \textbf{BAPMw/oA}) without using augmented data. 
In Fig~\ref{figDataAug_on_Seg}, we visualize a typical MRI with motion artifacts and its segmentation maps generated by FSL, BAPMw/oA, and BAPM. 
It can be observed from Fig.~\ref{figDataAug_on_Seg} that the segmentation maps of three tissues produced by BAPM and BAPMw/oA are generally better than those of FSL, suggesting the effectiveness of our methods in suppressing noisy information in original MRIs to some extent. 
Besides, as illustrated in red boxes in Fig.~\ref{figDataAug_on_Seg}, BAPMw/oA and FSL misidentify some CSF sub-regions as background, while BAPM can generate very good segmentation results in these regions. 
It implies that the proposed data augmentation strategy used in BAPM helps increase the diversity of input MR images, thus facilitating high-quality tissue segmentation. 

\subsection{Limitations and Future Work}
Several issues need to be considered in the future. 
\emph{First}, we use MRI reconstruction and tissue segmentation as two auxiliary tasks for modeling brain anatomy prior in this work, while other auxiliary tasks such as brain parcellation and MRI-to-CT translation can also be employed. 
As an interesting future work, we will explore other auxiliary tasks in the pretext model to more comprehensively capture brain anatomy priors. 
\emph{Second}, there exist data distribution differences between the source ADNI domain and two target domains, which may negatively affect the adaptability of our pre-trained encoder. 
Accordingly, we will utilize advanced harmonization methods~\cite{kamnitsas2017unsupervised,guan2021domain} to reduce such data heterogeneity between different studies. 
\emph{In addition}, we only use MRIs from ADNI for pretext model training currently. 
There are many other public brain MRI datasets (\eg, AIBL~\cite{ellis2009australian}, OASIS~\cite{marcus2007open}, SRPBS~\cite{tanaka2021multi}, and OpenBHB~\cite{dufumier2022openbhb}, and UK Biobank~\cite{allen2014uk}) that can be employed, which will also be our future work.

\section{Conclusion}
\label{S6}
In this paper, we develop a brain anatomy prior modeling (BAPM) framework to forecast clinical progression of cognitive impairment based on structural MRIs. 
Our BAPM can effectively learn a generalizable brain anatomy feature encoder by two auxiliary tasks (\ie, MRI reconstruction and brain tissue segmentation) on 9,344 public source MRIs without diagnostic label information. 
The pre-trained encoder can be transferred to different downstream prediction tasks on target data. 
We experimentally validate the BAPM on two CI-related studies with 448 subjects, with results suggesting that our method outperforms several state-of-the-art methods in MRI-based CI progression prediction. 
In addition, our pretext model can be flexibly applied to target MRIs for high-quality MR image  reconstruction and brain tissue segmentation.


\vspace{-4pt}
\bibliographystyle{model2-names.bst}
\biboptions{authoryear}
\bibliography{refs}

\end{document}